\documentclass{article}

\usepackage{color,ulem,oldlfont,amssymb,epsf,stmaryrd,epsfig}
\usepackage[a4paper,top=3cm,bottom=3cm,left=3cm,right=3cm]{geometry}

\newtheorem{definition}{Definition}

\def\lra{\longrightarrow}
\def\ra{\rightarrow}

\begin{document}

\title{Execution traces and reduction sequences}
\author{Gilles Dowek\thanks{Inria and 
\'Ecole normale sup\'erieure de Paris-Saclay,
61, avenue du Pr\'esident Wilson,
94235 Cachan Cedex, France,
{\tt gilles.dowek@ens-paris-saclay.fr}.}}
\date{}
\maketitle

\thispagestyle{empty}

\section{Introduction}

Y.~Gurevich \cite{Gurevich} proposes a definition of the notion of
algorithm as a set of execution traces.  This definition permits to
introduce the notion of algorithm before that of program, and
independently of any specific programming language.
It also permits to distinguish
\begin{itemize} 
\item what a program is: a text, 
\item what a program does: an algorithm, 
\item what a program computes: a function mapping input values to output
values.
\end{itemize}

In this note, we defend that this notion of algorithm as a set of
execution traces is somewhat independent of the notion of abstract
state machine---although it fits very well with it. It can be
reformulated in the more general framework of small step operational
semantics \cite{Plotkin}.

Reformulating this idea in the context of small step operational
semantics permits to define a notion of execution trace, not only for
imperative programs, but for also for functional programs, including
higher-order ones, shedding new light on the controversy on the
definition of the notion of algorithm as sets of execution traces and
as recursive equations \cite{Moschovakis,BlassGurevich}.

\section{Algorithms as sets of execution traces}

\subsection{Sets of execution traces}
\label{sets}

Consider an algorithm that computes the largest odd divisor of a
natural number by dividing it by $2$ until an odd number is reached.

This algorithm can be defined as the set of traces
$$\langle  12 \rangle, \langle  6 \rangle, \langle  3 \rangle$$
$$\langle  8 \rangle, \langle  4 \rangle, \langle  2 \rangle, \langle  1 \rangle$$
$$\langle  7 \rangle$$
$$...$$

The algorithms defined in this way need not be
deterministic. For instance, the non deterministic algorithm
computing 
the largest odd divisor of a
natural number by dividing it by any divisor that is a power of $2$ 
until an odd number is reached
can be defined as the set of sequences
$$\langle  12 \rangle, \langle  3 \rangle$$
$$\langle  12 \rangle, \langle  6 \rangle, \langle  3 \rangle$$
$$\langle  8 \rangle, \langle  4 \rangle, \langle  1 \rangle$$
$$\langle  8 \rangle, \langle  4 \rangle, \langle  2 \rangle, \langle  1 \rangle$$
$$\langle  7 \rangle$$
$$...$$

\subsection{The identity of two algorithms}

This definition also yields a notion of identity of two algorithms
\cite{BlassDershowitzGurevich}. For instance the two algorithms of
Section \ref{sets} computing the largest odd divisor of a natural
number are different, and are different from that containing the 
sequences
$$\langle 12 \rangle, \langle 12, 1 \rangle, \langle 12, 2 \rangle, 
\langle 12, 4 \rangle, \langle 3 , 4 \rangle, \langle 3 \rangle$$
$$\langle 8 \rangle, \langle 8, 1 \rangle, \langle 8, 2 \rangle, 
\langle 8, 4 \rangle, \langle 8, 8 \rangle, \langle 1 , 8 \rangle, 
\langle 1 \rangle$$
$$\langle 7 \rangle, \langle 7, 1 \rangle, \langle 7, 1 \rangle, 
\langle 7 \rangle$$
$$...$$ 

\section{From small step operational semantics to execution traces}

Not all sets of sequences are algorithms.  For instance, the set
containing the sequences of natural numbers $\langle \langle n_1 \rangle, \langle n_2 \rangle \rangle$,
such that $n_2 = 1$ if $n_1$ is the G\"odel number of a terminating
program and $n_2 = 0$ otherwise is not, because the notion of execution
trace presuposes a notion of effectivity.

This notion of effectivity can be defined in the framework of small
step operational semantics.

\subsection{Small step operational semantics for functional languages}

Small step operational semantics has first been defined for functional
programming languages.

For instance, to define the result of the execution of the program {\tt fun x $\ra$
x * x + 1} on the value $7$, we first build the term {\tt (fun x
$\ra$ x * x + 1) 7} that contains both the program and the input
value and then reduce it, step by step, to the irreducible term {\tt
50}, with rewrite rules defining the semantics of the language.

For instance, with the $\beta$-reduction rule and obvious rules for
the addition and multiplication, we have the following reduction
sequence
\begin{center}
{\tt (fun x $\ra$ x * x + 1) 7} $\lra$ {\tt 7 * 7 + 1}
$\lra$ {\tt 49 + 1} $\lra$ {\tt 50}
\end{center}

Again, not all sets of sequences are sets of reduction sequences. 
A set of sequences is a set of reduction sequences, if there 
exists a computable function, computing the next element of a sequence,
from the previous one---and, in the non deterministic case, the finite
set of possible next elements from the previous one.

\begin{definition}[Set of reduction sequences]
A set $R$ of sequences is a set of reduction sequences if
there exists a set $I$ of initial states and computable function {\it step}
such that $R$ is the set of all sequences
$s_1, ..., s_n$ such that $s_1$ is an element of $I$,
$s_n$ is such that $\mbox{\it step}(s_n) = \varnothing$ and 
for all $i$, $s_{i+1} \in \mbox{\it step}(s_i)$.
\end{definition}

The computable function {\it step} defines the granularity of the
reduction. For instance, a $\beta$-reduction step can be considered as
atomic, or as a sequence of steps whose length depends on the size of
the body of the abstraction, like in the calculus of explicit
substitutions \cite{ACCL}. In the same way, the assignment of a variable
can be considered as an atomic step or as a sequence of steps whose
length depends, logarithmically, on the size of the memory.

This choice of a granularity affects the length of the reduction
sequences, hence the complexity of the algorithms. The complexity of
algorithms is thus relative to the choice of such as function {\it
step}.

\subsection{Small step operational semantics for imperative languages}

\begin{figure}
\framebox{\parbox{\textwidth}{
\begin{center}
{\tt $\langle$skip;r,$\rho \rangle \lra \langle$r,$\rho \rangle$}
\end{center}

\begin{center}
{\tt $\langle$x := t;r,$\rho \rangle \lra \langle$r, 
$\rho + ($x $= \llbracket$t$\rrbracket_{\rho}) \rangle$}
\end{center}

\begin{center}
{\tt $\langle$(p;q);r,$\rho \rangle \lra \langle$p;(q;r),$\rho \rangle$}
\end{center}

If {\tt $\llbracket$t$\rrbracket_{\rho} =$ true} then 

\begin{center}
{\tt $\langle$if t then p else q;r,$\rho \rangle \lra \langle$p;r,$\rho \rangle$}
\end{center}

If {\tt $\llbracket$t$\rrbracket_{\rho} =$ false} then 

\begin{center}
{\tt $\langle$if t then p else q;r,$\rho \rangle \lra \langle$q;r,$\rho \rangle$}
\end{center}

\begin{center}
{\tt $\langle$while t do p;r,$\rho \rangle \lra \langle$(if t then p;while t do p else skip);r,$\rho \rangle$}
\end{center}
}}
\caption{Small step operational semantics of a simple imperative language \label{imp}}
\end{figure}

An imperative program, such as 
\begin{center}
{\tt while even(x) do x := x / 2}
\end{center}
cannot, alone, be reduced by small step operational semantics rules.
But, the ordered pair formed with this program and a state such as
{\tt $\langle$x $= 12 \rangle$} can be reduced to the pair formed with
the empty program and the state {\tt $\langle$x $= 3 \rangle$}.  The
small step operational semantics of a simple imperative language is
given in Figure \ref{imp}, considering that {\tt r} may be empty, in
which case {\tt p;r} is just {\tt p}. Irreducible terms are those
whose program is empty.  For instance, the reduction sequence of this
program in the state {\tt $\langle$x $= 12 \rangle$} is the sequence

{\tt $\langle$while even(x) do x := x / 2$, \langle$x $ = 12 \rangle \rangle$}

{\tt $\lra \langle$if even(x) then x := x /2; while even(x) do x := x
  / 2 else skip$, \langle$x $= 12 \rangle \rangle$}

{\tt $\lra \langle$x := x /2; while even(x) do x := x / 2$, \langle$x
  $= 12 \rangle \rangle$}

{\tt $\lra \langle$while even(x) do x := x / 2$, \langle$x
  $= 6 \rangle \rangle$}

{\tt $\lra \langle$if even(x) then x := x /2; while even(x) do x := x
  / 2 else skip$, \langle$x $= 6 \rangle \rangle$}

{\tt $\lra \langle$x := x /2; while even(x) do x := x / 2$, \langle$x
  $= 6 \rangle \rangle$}

{\tt $\lra \langle$while even(x) do x := x / 2$, \langle$x
  $ = 3 \rangle \rangle$}

{\tt $\lra \langle$if even(x) then x := x /2; while even(x) do x := x
  / 2 else skip$, \langle$x $= 3 \rangle \rangle$}

{\tt $\lra \langle$skip$, \langle$x $ = 3 \rangle \rangle$}

{\tt $\lra \langle, \langle$x $= 3 \rangle \rangle$}

\subsection{From reduction sequences to execution traces}

The notion of execution trace of an imperative program can easily be
defined from such a reduction sequence: if the reduction sequence is
\begin{center}
{\tt $\langle$p$_{\tt 1}, \rho_1 \rangle \lra \langle$p$_{\tt 2}, \rho_2 \rangle \lra$~...~$\lra \langle$p$_{\tt n}, \rho_n \rangle$}
\end{center}

then the execution trace of {\tt p$_{\tt 1}$} in $\rho_1$ is the sequence
obtained by dropping the program in each pair and errasing variables
names in the states.

For instance, the execution trace of the program 
\begin{center}
{\tt while even(x) do x := x / 2}
\end{center}
in the state $\langle {\tt x} = 12 \rangle$ is
$$\langle 12 \rangle, \langle 12 \rangle, \langle 12 \rangle, \langle 6
\rangle, \langle 6 \rangle, \langle 6 \rangle, \langle 3 \rangle,
\langle 3 \rangle, \langle 3 \rangle, \langle 3 \rangle$$

More generally, a execution trace is any projection of a reduction
sequence.  This rules out algorithms computing non computable
functions and guarantees effectiveness.

\begin{definition}[Projection]
Let {\it proj} be a computable function.  If $s$ is a sequence $s_1,
..., s_n$ we write $\mbox{proj}(s)$ for the pointwise application of
the function {\it proj} to the elements of this sequence: $\mbox{\it
  proj}(s_1), ..., \mbox{\it proj}(s_n)$.

If $R$ is a set of
sequences, we write $\mbox{\it proj}(R)$ for the pointwise application
of the function {\it proj} to the elements of $R$.

We say that a set $A$ is a projection of a set $R$ if there exists a
computable function {\it proj} such that $A = \mbox{proj}(R)$.
\end{definition}

\subsection{Speeding up}

The set of execution traces
$$\langle 12 \rangle, \langle 12 \rangle, \langle 12 \rangle, \langle 6
\rangle, \langle 6 \rangle, \langle 6 \rangle, \langle 3 \rangle,
\langle 3 \rangle, \langle 3 \rangle, \langle 3 \rangle$$
$$\langle 8 \rangle, \langle 8 \rangle, \langle 8 \rangle,
\langle 4 \rangle, \langle 4 \rangle, \langle 4 \rangle,
\langle 2 \rangle, \langle 2 \rangle, \langle 2 \rangle,
\langle 1 \rangle, \langle 1 \rangle, \langle 1 \rangle, \langle 1 \rangle$$
$$\langle 7 \rangle, \langle 7 \rangle, \langle 7 \rangle, \langle 7 \rangle$$
is different from the set of execution traces of Section \ref{sets}.

But, the sequence 
$$\langle 12 \rangle, \langle 6 \rangle, \langle 3 \rangle$$
is a subsequence of the sequence 
$$\langle 12 \rangle, \langle 12 \rangle, \langle 12 \rangle, \langle 6
\rangle, \langle 6 \rangle, \langle 6 \rangle, \langle 3 \rangle,
\langle 3 \rangle, \langle 3 \rangle, \langle 3 \rangle$$
and the
number of consecutive omitted steps is bounded by $3$.

This leads to define a notion of speed up of a set of sequences.

\begin{definition}[Speed up]
A set of sequences $B$ is a speed up of a set of sequences $A$ if
there exists a number $k$ such that each sequence of $B$ is a subsequence
of one of $A$, containing the same first and last elements and such that 
the number of consecutive omitted elements is bounded by $k$.
\end{definition}

Effectiveness is preserved by a speed up, and complexity also
as the length of execution traces is at most multiplied or divided by
the constant $k$.

We can now give a formal definition of the notion of algorithm.

\begin{definition}[Algorithm]
An algorithm is a set of sequences that is a speed up of a projection of
a set of reduction sequences.
\end{definition}

For instance, the initial set containing all the pairs formed with
the program 
\begin{center}
{\tt while even(x) do x := x / 2}
\end{center} 
and a state {\tt $\langle$x $= n \rangle$} for some natural number
$n$, and the {\it step} function defined in Figure \ref{imp} define a
set of reduction sequence. The projection function {\it proj} erasing
the program and the variable names define an algorithm containing
the traces 
$$\langle 12 \rangle, \langle 12 \rangle, \langle 12 \rangle, \langle 6
\rangle, \langle 6 \rangle, \langle 6 \rangle, \langle 3 \rangle,
\langle 3 \rangle, \langle 3 \rangle, \langle 3 \rangle$$
$$\langle 8 \rangle, \langle 8 \rangle, \langle 8 \rangle,
\langle 4 \rangle, \langle 4 \rangle, \langle 4 \rangle,
\langle 2 \rangle, \langle 2 \rangle, \langle 2 \rangle,
\langle 1 \rangle, \langle 1 \rangle, \langle 1 \rangle, \langle 1 \rangle$$
$$\langle 7 \rangle, \langle 7 \rangle, \langle 7 \rangle, \langle 7 \rangle$$
that has a a speed up the algorithm containing the traces 
$$\langle  12 \rangle, \langle  6 \rangle, \langle  3 \rangle$$
$$\langle  8 \rangle, \langle  4 \rangle, \langle  2 \rangle, \langle  1 \rangle$$
$$\langle  7 \rangle$$
$$...$$

So, both algorithms are expressed by the program 
\begin{center}
{\tt while even(x) do x := x / 2}
\end{center} 
Thus, this program does not express a single algorithm, but several depending
on the granularity, defined by the {\it step} function and by the speed up,
at which this program is observed.

\section{Functional languages}

The very idea of functional programming is that programs do not do
anything, but just compute something. So it is often assumed that
the notion of execution trace---what the program does---makes
no sense for functional languages.

The absence of a notion of execution trace for functional programs is
probably at the root of the controversy \cite{BlassGurevich}
on the definition of the notion of algorithm as sets of execution
traces \cite{Gurevich} and as recursive equations \cite{Moschovakis}.

But, as we have seen, the functional languages have a small step
operational semantics, and, as we shall see, this small step
operational semantics permits to extend the notion of execution trace
to functional programs.

\subsection{Recursive equations}
\label{recursive}

Consider, for instance the functional program, that computes the largest 
odd factor of a natural number, defined by the recursive equation
\begin{center}
{\tt f(x) = if(even(x),f(/(x,2)),x)}
\end{center}
or the rewrite rule 
\begin{center}
{\tt f(x) $\lra$ if(even(x),f(/(x,2)),x)}
\end{center}
with obvious rules to compute parity, division and test.

Then, the term {\tt f(12)} reduces to the term {\tt 3}. But, in small 
step operational semantics, the reduction sequence cannot just be

{\tt f(12)}

{\tt $\lra$ if(even(12),f(/(12,2)),12)}

{\tt $\lra$ if(true,f(/(12,2)),12)}

{\tt $\lra$ f(/(12,2))}

{\tt $\lra$ f(6)}

{\tt $\lra$ if(even(6),f(/(6,2)),6)}

{\tt $\lra$ if(true,f(/(6,2)),6)}

{\tt $\lra$ f(/(6,2))}

{\tt $\lra$ f(3)}

{\tt $\lra$ if(even(3),f(/(3,2)),3)}

{\tt $\lra$ if(false,f(/(3,2)),3)}

{\tt $\lra$ 3}

\noindent 
because the fact that {\tt f(12)} reduces to {\tt
  if(even(12),f(/(12,2)),12)} is not part of the semantics of the
language, and depends on the program.  So, like for imperative
languages, we need to introduce a pair formed with a program---a set
of recursive equations---and a term, and if we write {\tt p} for the
set of recursive equations above, the reduction sequence is then

{\tt $\langle$p,f(12)$\rangle$}

{\tt $\lra \langle$p,if(even(12),f(/(12,2)),12)$\rangle$}

{\tt $\lra \langle$p,if(true,f(/(12,2)),12)$\rangle$}

{\tt $\lra \langle$p,f(/(12,2))$\rangle$}

{\tt $\lra$} ...

\noindent
Thus, the sequence of terms 

{\tt f(12)}

$\lra$ 
{\tt if(even(12),f(/(12,2)),12)}

$\lra$ 
{\tt if(true,f(/(12,2)),12)}

$\lra$ 
{\tt f(/(12,2))}

$\lra$ ...

\noindent 
that contains program names, {\tt f}, {\tt /}, ... but no programs, is
the analog of sequence of states in imperative languages.

In the definition of a trace for imperative programs we have dropped
variables names. Exactly in the same way, we can drop program names and 
get the execution trace

\medskip

{\tt <fun>(12)},

{\tt <fun>(<fun>(12),<fun>(<fun>(12,2)),12)},

{\tt <fun>(true,<fun>(<fun>(12,2)),12)},

{\tt <fun>(<fun>(12,2))},

{\tt <fun>(6)},

{\tt <fun>(<fun>(6),<fun>(<fun>(6,2)),6)},

{\tt <fun>(true,<fun>(<fun>(6,2)),6)},

{\tt <fun>(<fun>(6,2))},

{\tt <fun>(3)},

{\tt <fun>(<fun>(3),<fun>(<fun>(3,2)),3)},

{\tt <fun>(false,<fun>(<fun>(3,2)),3)},

{\tt 3}

\subsection{Lambda-calculus style languages}

When functional programs are defined with recursive equations or with
rewrite rules, it is possible to separate the state from the program,
and thus to define a notion of execution trace. But, in other
functional languages, such as the lambda-calculus or the rho-calculus
\cite{rho}, the program and the data are less easy to separate.

We consider an extension of the lambda-calculus where the {\tt fun}
operation is recursive, called {\tt fixfun} in \cite{DowekLevy}, and 
the reduction rule 
\begin{center}
{\tt (fun f x $\ra$ t) u $\lra$ (u/x,fun f x $\ra$ t/f)t}
\end{center}
with obvious rules to compute parity, division and test.

Then, the term {\tt (fun f x $\ra$ (if (even x) (f (/ x 2)) x)) 12}
reduces to the term {\tt 3}. 

{\tt (fun f x $\ra$ (if (even x) (f (/ x 2)) x)) 12}

{\tt $\lra$ ((fun f x $\ra$ (if (even x) (f (/ x 2)) x)) 12)}

{\tt $\lra$ (if (even 12) ((fun f x $\ra$ (if (even x) (f (/ x 2)) x)) (/ 12 2)) 12)}

{\tt $\lra$ (if true ((fun f x $\ra$ (if (even x) (f (/ x 2)) x)) (/ 12 2)) 12)}

{\tt $\lra$ ((fun f x $\ra$ (if (even x) (f (/ x 2)) x)) (/ 12 2)))}

{\tt $\lra$ ((fun f x $\ra$ (if (even x) (f (/ x 2)) x)) 6)}

{\tt $\lra$ (if (even 6) ((fun f x $\ra$ (if (even x) (f (/ x 2)) x)) (/ 6 2)) 6)}

{\tt $\lra$ (if true ((fun f x $\ra$ (if (even x) (f (/ x 2)) x)) (/ 6 2)) 6)}

{\tt $\lra$ ((fun f x $\ra$ (if (even x) (f (/ x 2)) x)) (/ 6 2))}

{\tt $\lra$ ((fun f x $\ra$ (if (even x) (f (/ x 2)) x)) 3)}

{\tt $\lra$ (if (even 3) ((fun f x $\ra$ (if (even x) (f (/ x 2)) x)) (/ 3 2)) 3)}

{\tt $\lra$ (if false ((fun f x $\ra$ (if (even x) (f (/ x 2)) x)) (/ 3 2)) 3)}

{\tt $\lra$ 3}

\noindent 
To separate the programs from the data, we replace every abstraction 
by the constant {\tt <fun>} and get the execution trace

{\tt (<fun> 12)},

{\tt (<fun> (<fun> 12) (<fun> (<fun> 12 2)) 12)},

{\tt (<fun> true (<fun> (<fun> 12 2)) 12)},

{\tt (<fun> (<fun> 12 2)))},

{\tt (<fun> 6)},

{\tt (<fun> (<fun> 6) (<fun> (<fun> 6 2)) 6)},

{\tt (<fun> true (<fun> (<fun> 6 2)) 6)},

{\tt (<fun> (<fun> 6 2))},

{\tt (<fun> 3)},

{\tt (<fun> (<fun> 3) (<fun> (<fun> 3 2)) 3)},

{\tt (<fun> false (<fun> (<fun> 3 2)) 3)},

{\tt 3}

\noindent
Note that, with minor notation changes, this execution trace is the
same as that in the the language of Section \ref{recursive}. 

\section{When do two functional programs express the same
  algorithm?}

This notion of execution trace permits to define algorithms as a set
of execution traces and define when two functional programs execute
the same algorithm.  For this definition, although syntactically
different, the two programs computing the identity function on
Booleans

\begin{center}
{\tt f(false) = false}

{\tt f(true) = true}
\end{center}
and 
\begin{center}
{\tt f(x) = x}
\end{center}
express the same algorithm
\begin{center}
{\tt <fun>(false), false} 

{\tt <fun>(true), true}
\end{center}

\section*{Acknowledgements}

Many thanks to the members of the working group Tarmac for
enlightening discussions.

\bibliographystyle{plain}
\bibliography{traces}

\end{document}